\documentclass[rnote,traditabstract]{aa}

\usepackage{amssymb}
\usepackage{amsmath}
\usepackage{txfonts}
\usepackage{graphicx}
\usepackage[usenames]{xcolor}
\usepackage[bookmarks, colorlinks, breaklinks]{hyperref}
\hypersetup{linkcolor=blue,citecolor=blue,filecolor=black,urlcolor=blue} 
\usepackage{natbib}
\bibpunct{(}{)}{;}{a}{}{,} 
\begin{document}

\title{Using hierarchical octrees \\ in Monte Carlo radiative
  transfer simulations}

\titlerunning{Octrees in Monte Carlo radiative transfer}

\author{
W. Saftly\inst{\ref{UGent}}, 
P. Camps\inst{\ref{UGent}},   
M. Baes\inst{\ref{UGent}}, 
K. D. Gordon\inst{\ref{STScI},\ref{UGent}},
S. Vandewoude\inst{\ref{UGent}}, 
A. Rahimi\inst{\ref{CAS}},
M. Stalevski\inst{\ref{UGent},\ref{Belgrade},\ref{INI}}
}

\authorrunning{W. Saftly et al.}

\institute{ 
  Sterrenkundig Observatorium, Universiteit Gent, Krijgslaan
  281, B-9000 Gent, Belgium\label{UGent} 
  \and
  Space Telescope Science Institute, 3700 San Martin Drive, Baltimore,
  MD 21218, USA\label{STScI}  
  \and
   National Astronomical Observatories, Chinese Academy of Sciences, 
   Beijing 100012, China\label{CAS}
  \and 
  Astronomical Observatory, Volgina 7, 11060 Belgrade,
  Serbia\label{Belgrade}
  \and
  Isaac Newton Institute of Chile, Yugoslavia Branch, Volgina 7, 11060
  Belgrade, Serbia\label{INI}
}

\date{\today}

\abstract{A crucial aspect of 3D Monte Carlo radiative transfer is the
  choice of the spatial grid used to partition the dusty medium.  We
  critically investigate the use of octree grids in Monte Carlo dust
  radiative transfer, with two different octree construction
  algorithms (regular and barycentric subdivision) and three different
  octree traversal algorithms (top-down, neighbour list, and the
  bookkeeping method).  In general, regular octree grids need higher
  levels of subdivision compared to the barycentric grids for a fixed
  maximum cell mass threshold criterion. The total number of grid
  cells, however, depends on the geometry of the model. 
Surprisingly, regular octree grid simulations turn out to be 10 to
  20\% more efficient in run time than the barycentric grid
  simulations, even for those cases where the latter contain fewer
  grid cells than the former. Furthermore, we find that storing
  neighbour lists for each cell in an octree, ordered according to
  decreasing overlap area, is worth the additional memory and
  implementation overhead: using neighbour lists can cut down the grid
  traversal by 20\%  compared to the traditional top-down
    method. In conclusion, the combination of a regular node
  subdivision and the neighbour list method results in the most
  efficient octree structure for Monte Carlo radiative transfer
  simulations.}


\maketitle

\section{Introduction}

Cosmic dust is present in and around many astrophysical systems,
ranging from planetary and stellar atmospheres to the diffuse
interstellar medium in galaxies. Dust grains have a profound effect on
the radiation field, as they efficiently scatter, absorb, and re-emit
radiation from these sources. Radiative transfer calculations are
required if we want to understand the intrinsic properties of dusty
objects, or predict the observable properties of artificial systems
simulated using hydrodynamical simulations. In the past decade,
several codes have been developed that can handle the full dust
radiative transfer problem in a general 3D geometry; for a general
overview of 3D dust radiative transfer we refer to
\citet{SBG}. 

Virtually all 3D radiative transfer codes are based on the Monte Carlo
technique \citep[e.g.][]{2001ApJ...551..269G, 2003A&A...397..201J,
  2003CoPhC.150...99W, 2006MNRAS.372....2J, 2006A&A...459..797P,
  2011ApJS..196...22B} and \citep{2011A&A...536A..79R}.\footnote{Monte Carlo
  techniques have been applied to different 3D transport problems,
  including neutron, neutrino, UV ionizing radiation, and Ly$\alpha$
  radiation transport \citep[e.g.][]{2001MNRAS.324..381C,
    2004MNRAS.348.1337W, 2006A&A...460..397V, 2006ApJ...645..792T,
    2009ApJ...696..853L} and  \citep{2012ApJ...755..111A}. The focus of this
  Research Note is on dust radiative transfer, but the results are
  equally applicable to other Monte Carlo transport problems.}  In
Monte Carlo dust radiative transfer simulations, the dusty medium is
divided into a large number of tiny grid cells. Each of these cells is
typically characterized by a constant dust density, temperature,
radiation field, etc. Their size sets the effective resolution of the
simulation. On the one hand, it is useful to maximize the number of
cells, in order to obtain the highest resolution possible. On the
  other hand, the memory requirements and run time of the simulation
  scales as $\mathcal{O}(N^{1/3})$ to $\mathcal{O}(N)$ with $N$ the
  number of grid cells, which puts a limit to this number. The choice
of the grid structure is, therefor, a crucial ingredient of a modern 3D
radiative transfer simulation. Ideally, the size of the grid cells
should be linked to the dust mass, optical depth or temperature of the
dusty medium: the cells should be small where the dust density is high
or the radiation field shows a large gradient, and they can be bigger
where the dust density is low and the radiation field does not change
significantly.

The most popular candidate for such grid structures are hierarchical
octree grids, in which the 3D space is partitioned by recursively
subdividing it into eight subcubes. Octrees are widely used in all
areas of science and engineering, especially in computer graphics and
3D game engines \citep{Jackins1980}. In astronomy, they are most
popular in N-body and hydrodynamics codes \citep{1986Natur.324..446B,
  1987ApJS...64..715H, 1989ApJS...70..419H, 2002A&A...385..337T} and
  \citep{ 2005MNRAS.364.1105S}. Octree grid structures, or more general
hierarchical grid-in-grid structures, have now been implemented in
several Monte Carlo dust radiative transfer codes
\citep[e.g.][]{2001A&A...379..336K, 2003CoPhC.150...99W,
  2004MNRAS.350..565H, 2006MNRAS.372....2J, 2008A&A...490..461B,
  2006A&A...456....1N, 2011A&A...536A..79R, 2012ApJ...751...27H} and 
   \citep{2012A&A...544A..52L}, and UV ionization and Ly$\alpha$ transfer
codes \citep[e.g.][]{2006ApJ...645..792T, 2009ApJ...696..853L}.

In this Research Note we investigate two different aspects of the use
of octree grid structures in 3D Monte Carlo radiative transfer
simulations. The first aspect concerns the construction of the grid,
and more specifically the way dust cells are subdivided. All radiative
transfer codes with octree dust grids subdivide the cells in a regular
way, whereas we investigate whether a barycentric subdivision is more
efficient. Second, we concentrate on the traversal of photon packages
through the dust grid. Most Monte Carlo codes use a simple top-down
method to traverse the octree. Other bottom-up octree traversal
algorithms have been developed in the field of computer graphics and
it remains to be investigated whether these might be more efficient in
the context of radiative transfer simulations. In
Sect.~{\ref{OctreeAlgorithms.sec}} we present the algorithms
proposed for the construction of the octree and the traversal of
photon packages through the octree. In Sect.~{\ref{TestModels.sec}}
we present three different test models on which we test the different
algorithms.  In Sect.~{\ref{Results.sec}} we present the results of
our simulations, and in Sect.~{\ref{Conclusion.sec}} we present the conclusion.

\section{Octree algorithms}
\label{OctreeAlgorithms.sec}

\subsection{Construction of the octree grid}

We have implemented an octree dust grid structure in the SKIRT Monte
Carlo radiative transfer code \citep{2003MNRAS.343.1081B,
  2011ApJS..196...22B} based on a dust mass threshold, in the sense
that each dust cell (i.e.\ each leaf node in the tree) can contain at
most a fraction $\delta_{\text{max}}$ of the total dust mass in the
system. The construction of the octree grid is done as it is in most
other radiative transfer codes with octrees
\citep{2001A&A...379..336K, 2003CoPhC.150...99W, 2004MNRAS.350..565H} and
   \citep{2008A&A...490..461B} and is very straightforward. We create a list of
nodes over which we run a loop. The first node in the list is a
cuboidal root node that encloses the entire dusty medium. While
running the loop, we test whether each node should be subdivided by
calculating the ratio $\delta = M/M_{\text{tot}}$, where $M$ is the
estimated dust mass in the node (determined using a Monte Carlo
integration) and $M_{\text{tot}}$ is the total dust mass in the
system. If $\delta<\delta_{\text{max}}$ there is no subdivision, the
node is a leaf node and hence an actual dust cell. On the other hand,
if $\delta>\delta_{\text{max}}$, we partition the node into eight
subnodes and append them at the end of the node list. In either case,
we subsequently move to the next node in the node list and repeat the
same test. This loop is repeated until the entire list of nodes is
finished.

\subsection{Octree subdivision}

Standard octree algorithms subdivide each node by taking the centre of
the node as the subdivision point, such that all eight subnodes of a
node have the same size (and in general, all nodes corresponding to
the same level of subdivision have exactly the same dimensions). We
denote this method as {\em{regular}} subdivision.  One might wonder
whether this is the ideal situation. In a node with a large density
gradient, applying a regular subdivision might result in a set of
eight subnodes in which nearly all the dust mass is concentrated in
only one of the subnodes, which might lead to frequent nested
subdivision. An alternative could be to choose the division point of
the node such that the mass distribution is divided more or less equally 
over the eight subnodes. This can be achieved by choosing the
barycentre rather than the geometrical centre of the node as the
subdivision point. With a {\em{barycentric}} subdivision, one would
expect to be able to zoom in faster on high density regions. It is
straightforward to see that both the memory penalty and the
computational cost are minimal compared to the complete simulation
memory requirement and run time.

\subsection{Octree traversal}

During a Monte Carlo simulation, the life cycle of every photon
package implies the calculation of several random straight paths
through the dust grid.  This essentially comes down to the
determination of the ordered list of dust cells that the path
intersects and the physical distance covered in each individual
cell. As a Monte Carlo simulation typically follows many millions of
photon packages, it is obvious that this task should be implemented as
efficiently as possible.

The grid traversal is a loop that consists of a repetition of two
simple steps. The first step is to determine in which cell the initial
path position is located, and the next step is to follow the path
to the boundary of the cell while recording the covered
distance. While the second step is straightforward, the first step is
less so in an octree grid structure. Actually, the traversal of rays
in octrees has been a topic of investigation in the computer graphics
literature \citep[e.g.][]{Glassner1984, Amanatides1987, Samet1989,
  Samet1990, Agate1991, Revelles2000} and  \citep{Frisken2002}. Loosely based on
these methods, we have implemented three different methods to perform
octree traversal as appropriate for Monte Carlo radiative transfer
simulations.

The most straightforward method to traverse an octree grid is a
top-down method \citep{Glassner1984}. The idea of this method is
simple: once a photon package hits the boundary of a cell and we have
determined the new position (which is now in a yet unidentified cell),
we start from the root node and descend the tree. For every node along
the tree, we determine in which of its subnodes the new position is
located. This loop ends when the node is a leaf node, i.e.\ an actual
dust cell. This top-down method is very easy to implement, and it does
not depend on whether the subdivision is regular or barycentric. This
algorithm is implemented in most Monte Carlo radiative transfer codes
that operate on octree grids \citep[e.g.][]{2001A&A...379..336K,
  2008A&A...490..461B}.

\begin{figure*}
  \centering
  \includegraphics[width=0.33\textwidth]{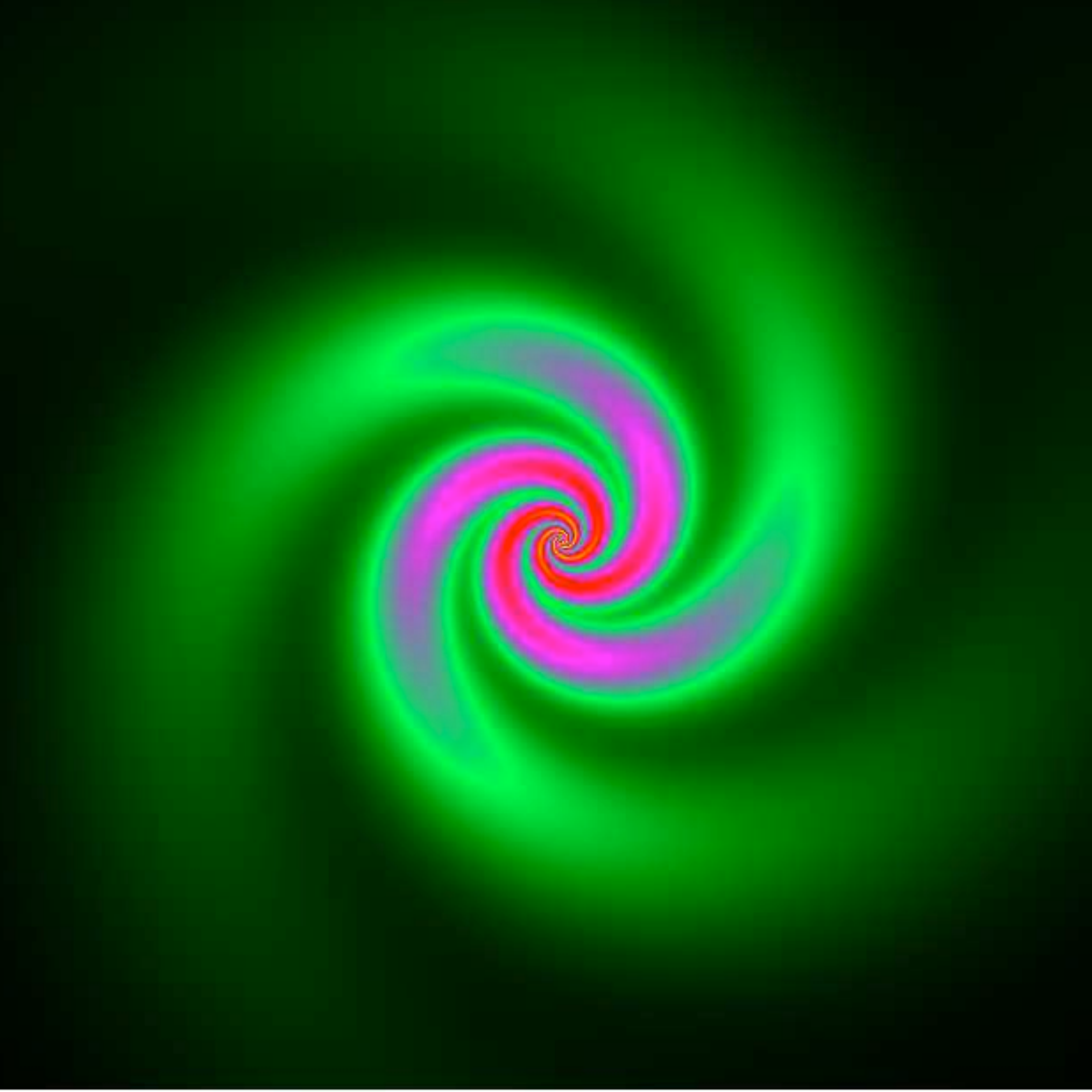}
  \includegraphics[width=0.33\textwidth]{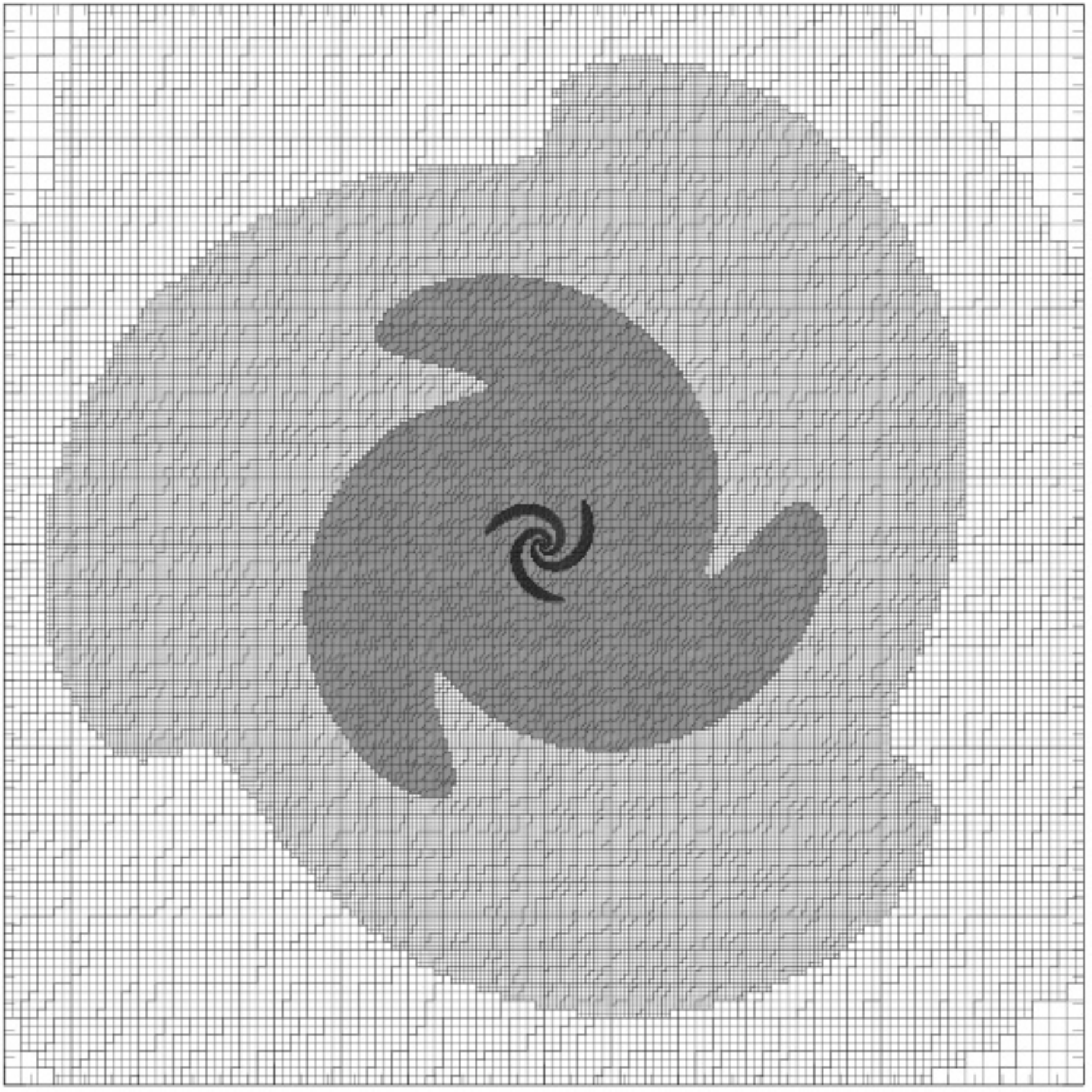}
  \includegraphics[width=0.33\textwidth]{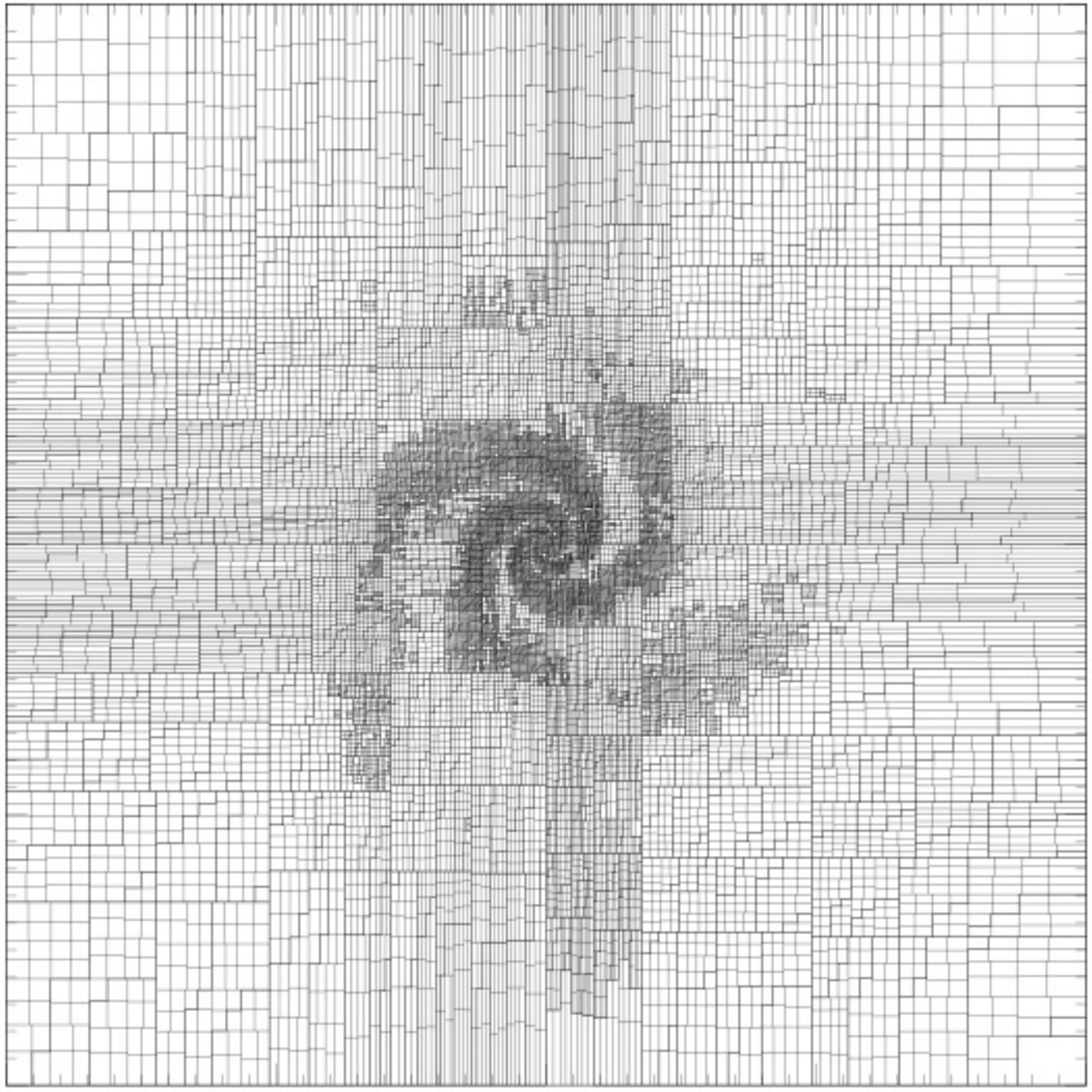}\\
  \includegraphics[width=0.33\textwidth]{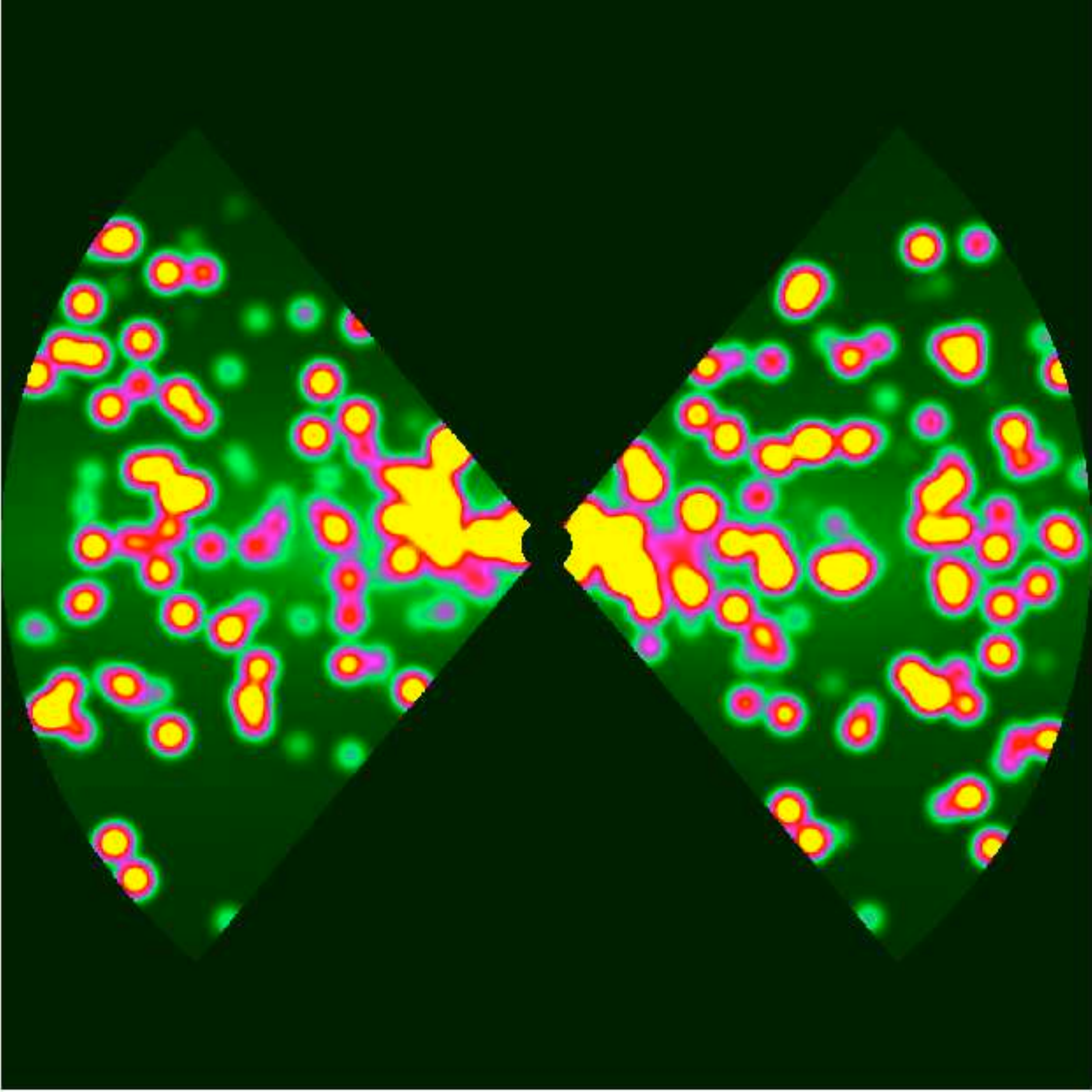}
  \includegraphics[width=0.33\textwidth]{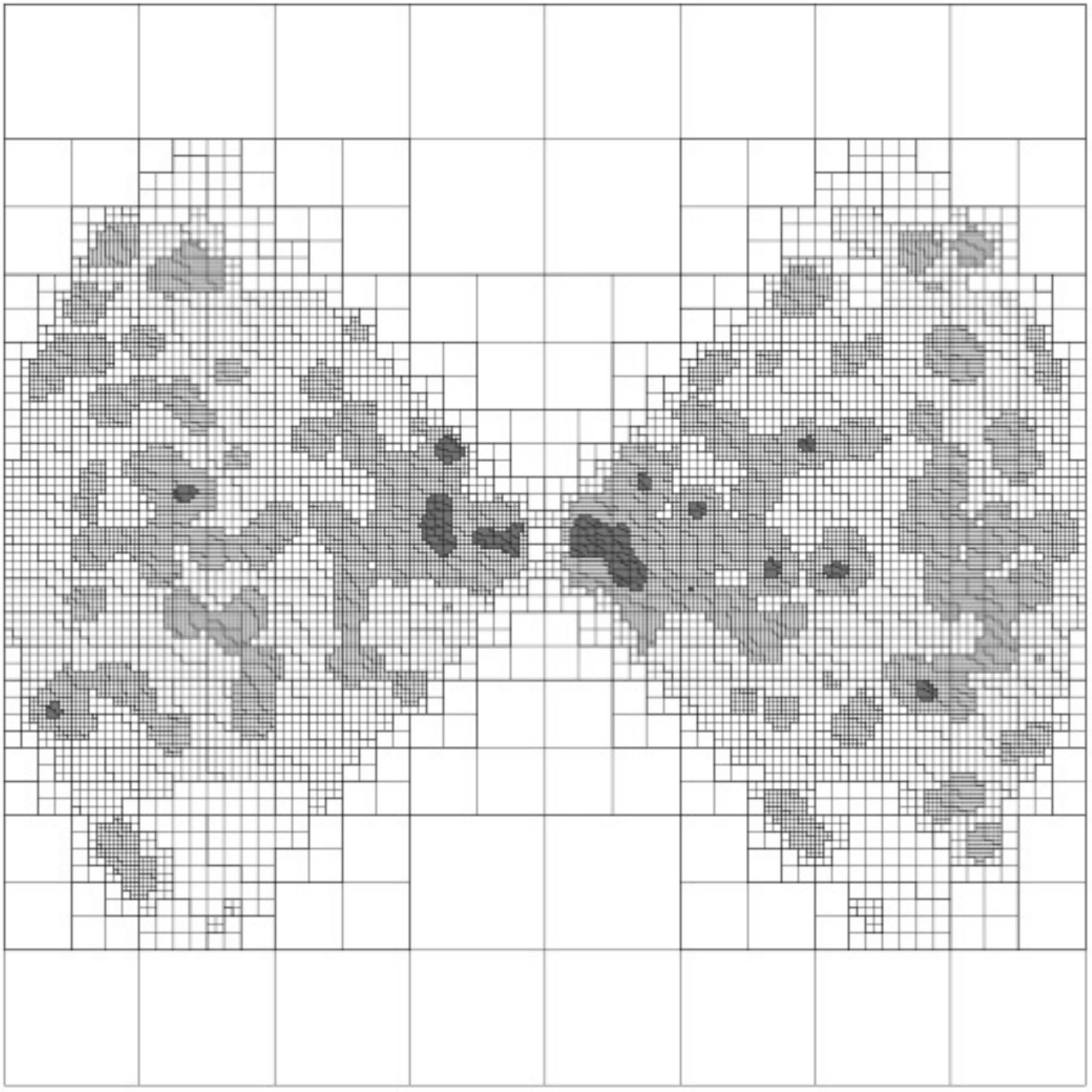}
  \includegraphics[width=0.33\textwidth]{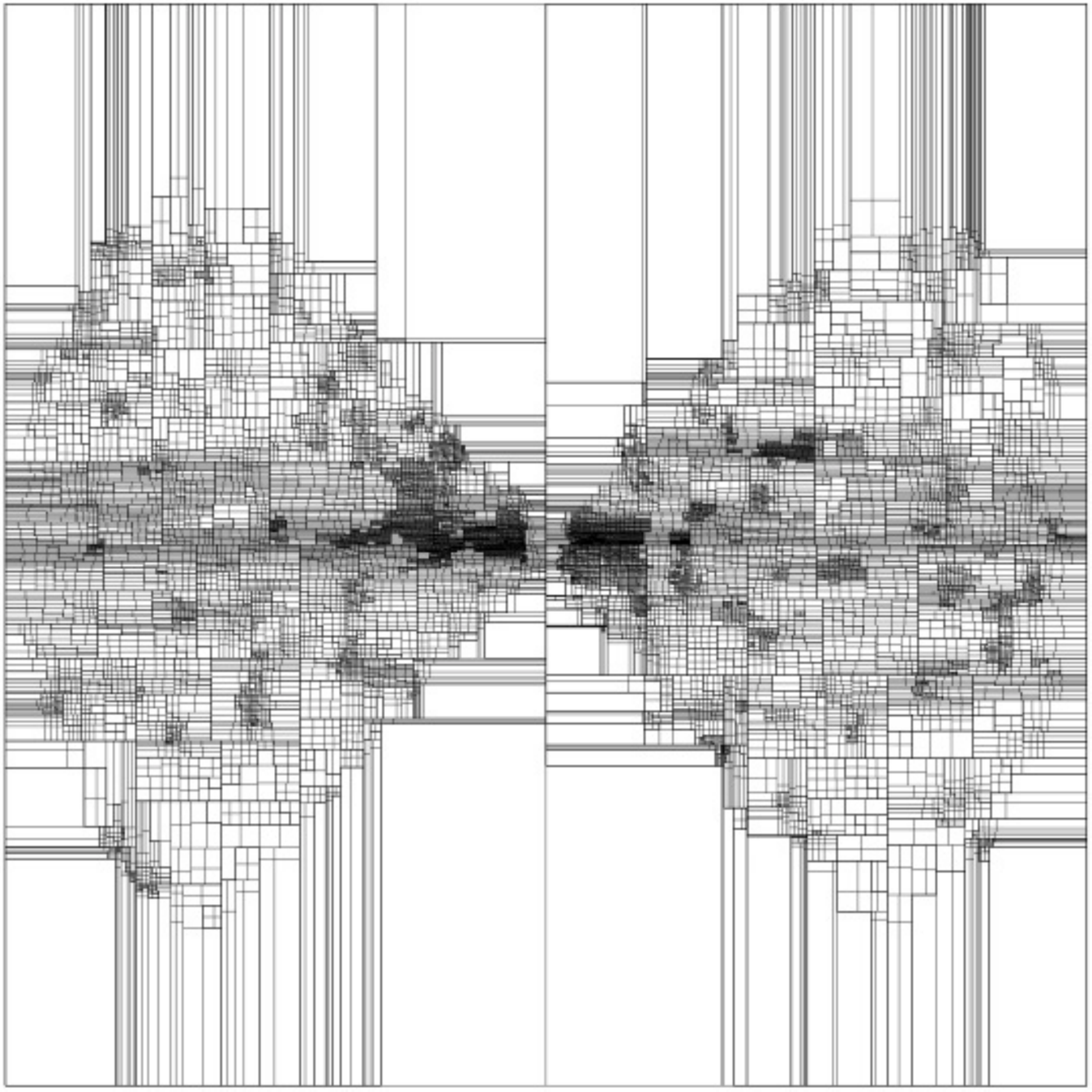}\\
  \includegraphics[width=0.33\textwidth]{SPHDensity.pdf} 
  \includegraphics[width=0.33\textwidth]{SPHRegular.pdf}
  \includegraphics[width=0.33\textwidth]{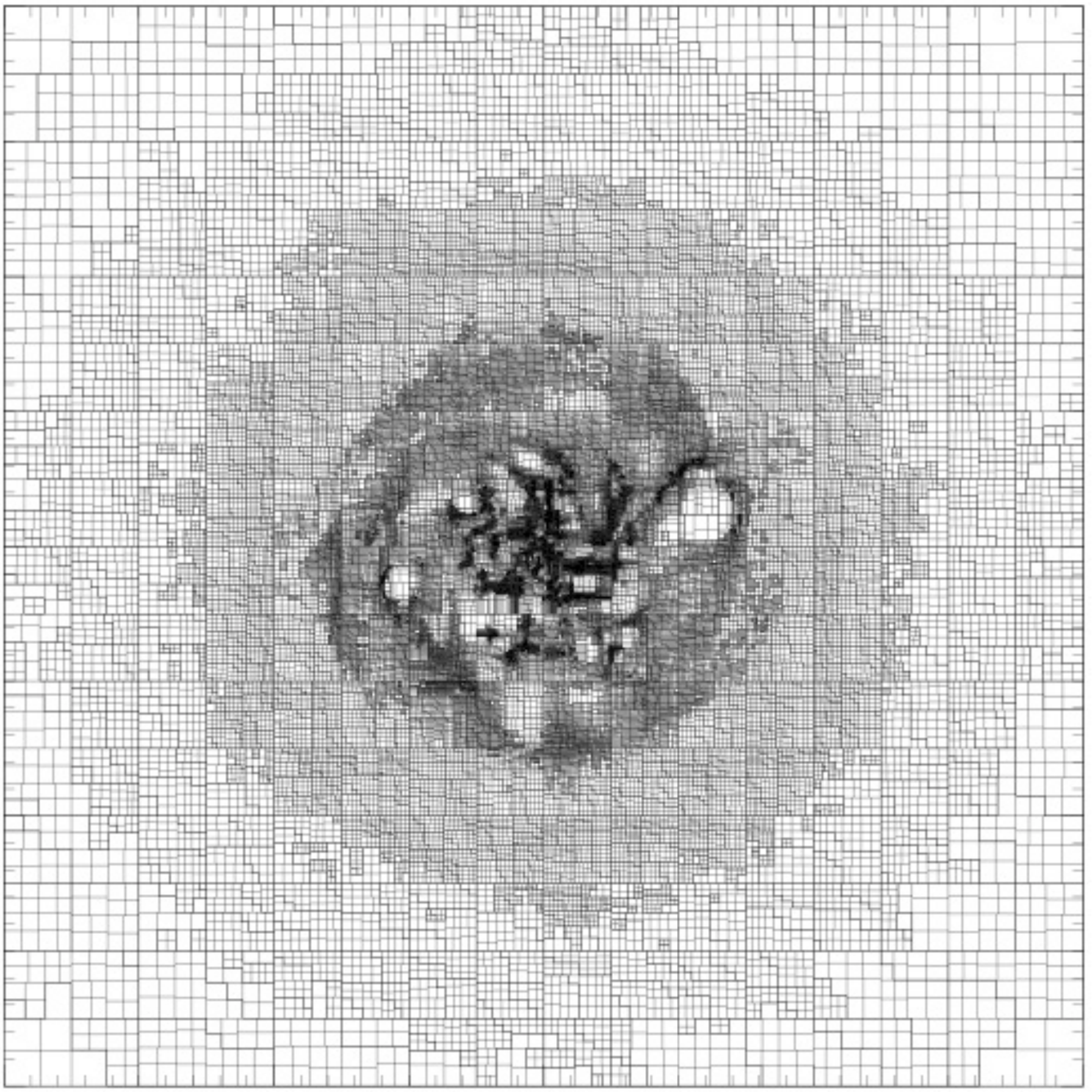}
  \caption{Illustration of the geometry and the octree grid structures
    for the three test models: the logarithmic spiral galaxy (top
    row), the AGN torus (middle row), and the galaxy from an SPH
    simulation (bottom row). In each row, the left column represents a
    cut through the dust density. For the two galaxy models, this cut
    corresponds to the $xy$ plane; for the AGN torus model it is a cut
    through the $xz$ plane. The central and right columns are cuts
    through the octree grids corresponding to the same planes, and
    correspond to the regular and barycentric subdivision recipes,
    respectively. The different shades of gray in the middle and
    right columns illustrate the density of the cells: darker gray
    means higher density. }
  \label{Models.fig}
\end{figure*}

An alternative method of traversing an octree makes use of a list of
the neighbouring cells of each node. This method was first advocated by
\citet{Samet1989, Samet1990} in the frame of computer graphics
ray-tracing. Instead of blindly looking for the cell that contains the
position of the photon package once it has traversed a cell wall, we
make use of the fact that the yet unknown cell is a neighbour of the
previous dust cell. In our implementation, every dust cell contains
six different lists (one for each cell wall) with pointers to the
neighbouring cells, ordered by decreasing area of overlap. If we know
through which wall the photon package has escaped the previous cell,
we just have to go down the corresponding list of neighbour cells. This
algorithm is straightforward and simple to implement, and again is
applicable both to regular and barycentric octree grids. The
additional memory usage of this method is usually not a problem: six
pointer lists need to be stored for each cell (one list for every cell
wall). For typical dust radiative transfer simulations that require
the storage of many properties per cell, this is a negligible
addition. Moreover, the neighbour lists can be created in a very
straightforward way during the construction of the octree grid, with
very limited computational overhead.  For the computations we
  discuss in this work, the computational overhead is less than 1 per
  cent of the simulation time; for realistic simulations, the relative
  overhead will be even much smaller.
 
Finally, we explored a third method of traversing an octree grid that
is inspired by the work of \citet{Frisken2002} and which we denote as
the bookkeeping method. One of the drawbacks of the top-down method is
that, for every passage through a wall, the search for the new cell
has to start from the root node. This is unfortunate, as we often know
in which branch of the tree the next position will be: indeed each
node in the octree has a place-awareness, i.e.\ the knowledge of its
position as a subnode with respect to its sibling nodes. If a
transition is internal, i.e.\ if the path crosses a cell wall that
does not leave the parent cell, we know that the next cell is one of
its siblings (or its children if the sibling is not a leaf node). A
slightly more difficult case occurs when the transition through a wall
is not an internal transition. One option could be to start the search
then from the root node as in the standard top-down method. An
alternative, which we have implemented, is to iteratively go up the
tree starting from the current node until the transition is an
internal transition, and then iteratively descend the tree until the
node is a leaf node.  Of the three methods presented, this method is
the most complex in terms of implementation. However, it is still very
manageable, it implies no memory overhead or additional computations,
and it is, again, applicable to regular as well as barycentric octree
grids.

\section{Test models}
\label{TestModels.sec}

To test these algorithms, we have considered three different
challenging test models. They were chosen to accommodate a variety of
possible geometries that can be encountered in realistic 3D Monte
Carlo radiative transfer simulations.

Our first model is an idealized model for a disc galaxy with a
three-armed logarithmic spiral structure. This model is completely
analytical and is inspired by the models used by
\citet{2000A&A...353..117M} and \citet{2012ApJ...746...70S}. Stars in
the model galaxy are distributed in two components: a
double-exponential disk with a spiral arm perturbation, and a
flattened S\'ersic bulge. The dust is only distributed in a
double-exponential disc; with a spiral perturbation.  For the
parameters of the stellar and dust distribution, we used typical
values applicable for spiral galaxies \citep{2002MNRAS.334..646K,
  2004A&A...414..905H} and  \citep{2012A&A...540A..52C}; for other parameters, we
were specifically inspired by the results obtained from radiative
transfer fits to nearby edge-on spiral galaxies
\citep{1999A&A...344..868X, 2007A&A...471..765B, 2010A&A...518L..39B} and  \citep{
  2011ApJ...741....6M}. An illustration of the dust density in the
central plane of the galaxy is shown on the top left panel of
Fig.~{\ref{Models.fig}}.

\begin{table*}
  \caption{Statistics of the different models and grids.}
  \label{Statistics.tab}
  \centering
  \begin{tabular}{|l|cc|cc|cc|}
    \hline
    \hline
& \multicolumn{2}{c|}{Spiral galaxy}  & \multicolumn{2}{c|}{AGN torus}& \multicolumn{2}{c|}{SPH galaxy}\\ 
& regular & barycentric & regular & barycentric & regular & barycentric \\
    \hline
    Top-down run time (s)
    & 878 & 975
    & 1171 & 1350 
    & 1142 & 1322 \\
    Neighbour list run time (s)
    & 713 & 800
    & 938 & 1104
    & 897 & 1082 \\
    Bookkeeping run time (s)
    & 893 & 1006 
    & 1168 & 1374
    & 1117 & 1351 \\
    Number of cells  
    & 3252264 & 3150295
    & 3050573 & 3383906 
    & 3315075 & 3373280 \\
   Average level of each cell 
    & 8.37 & 7.45
    & 7.99 & 7.56
    & 9.48 & 7.82 \\
    Average level of each cell crossed
    & 7.84 & 7.29
    & 7.67 & 7.19
    & 8.84 & 7.38 \\
   Average number of paths per photon package
   & 3.69 & 3.70
   & 1.85 & 1.99
   & 3.76 & 3.76 \\
   Average number of cells crossed per path
    & 94.8 & 102.5 
    & 126.9 & 139.7 
    & 124.8 & 150.3 \\
    Average number of neighbours per wall
    & 1.02 & 1.87 
    & 1.06 & 1.85 
    & 1.03 & 1.87 \\
    Average number of neighbours per wall crossed
    & 1.01 &  1.95 
    & 1.22  &  2.15 
    & 1.07  &  2.04  \\
    Average number of neighbours tested per wall crossed
    & 1.01 & 1.14 
    & 1.10 & 1.20 
    & 1.03 & 1.17 \\
    \hline
  \end{tabular}
\end{table*} 

Our second test model is a model for the central region of an active
galactic nucleus (AGN), and consists of a central, isotropic source
surrounded by an optically thick dust torus.  It is similar to the AGN
torus models presented by \citet{2012MNRAS.420.2756S}, in 
that it consists of a number of compact and optically thick clumps
embedded in a smooth interclump medium.  Contrary to the approach
adopted by \citet{2012MNRAS.420.2756S}, where the two-phase clumpy
medium was generated in a statistical way by applying a clumpiness
algorithm \citep{1990A&A...228..483B, 1996ApJ...463..681W} and  \citep{
  1998A&A...340..103W}, we now consider a torus model consisting of a
smooth distribution of dust to which we add 4,000 individual clumps
\citep[as in][]{2008A&A...490..461B}. An illustration of the dust
density in the $xz$ plane of the model is shown on the central left
panel of Fig.~{\ref{Models.fig}}.

The last test model we use is a completely numerical spiral galaxy
model created by means of a hydrodynamic simulation. The galaxy model
we consider is the 1 Gyr snapshot of model run number 6 from
\citet{2012MNRAS.422.2609R}, which represents an M33-sized late-type
spiral galaxy. It was run using the N-body/SPH code GCD+
\citep{2012MNRAS.tmp..115K} in a fixed dark matter halo, and includes
self-gravity, hydrodynamics, radiative cooling, star formation,
supernova feedback, metal enrichment and metal diffusion.  The
snapshot is characterized by 410,372 stellar particles and 189,628 gas
particles. For the calculation of the dust mass density distribution,
we follow the approach of \citet{2010MNRAS.403...17J} and assume that,
at every position in the galaxy, a fixed fraction of the metal content
is locked in dust grains. To obtain the dust density at a given
position, we interpolate the metal density over the SPH gas particles
and use a metal depletion in the dust grains of 40\%
\citep{1998ApJ...501..643D}. The resulting dust mass density in the
equatorial plane of the galaxy can be seen in the bottom left panel of
Fig.~{\ref{Models.fig}}. The holes and bubbles caused by the
feedback from supernovae and stellar winds are clearly visible.

\section{Results}
\label{Results.sec}

 In this section we compare the statistics on the number of cells in
the different models, and the efficiency of the path calculations
within the different grids. We stress that the different grid
structures for the six models (i.e.\ three geometries, each with
regular and barycentric octree subdivision) have been constructed
based on exactly the same criterion ($\delta_{\text{max}}=10^{-6}$).


Intuitively, one would expect that the barycentric subdivision is more
efficient in following the mass distribution. When a node is
subdivided with the barycentre as the subdivision point, the mass of
the parent node is redistributed more or less equally among the eight child
subnodes. This means that the subdivision needs to be repeated less
often in the barycentric case compared to the regular case. This is
confirmed in Row 5 of Table~{\ref{Statistics.tab}}, which
shows that the average level of a cell in the barycentric grids is
about 7.6, whereas it is about 8.6 for the regular grids. 

Surprisingly at first sight, this does not systematically
lead to a smaller total number of cells in the barycentric grids
compared to the equivalent regular octree grids (Row 4 of
Table~{\ref{Statistics.tab}}). For the spiral galaxy model, the
smoothest of the three test geometries, the barycentric grid contains
some 3\% fewer cells than the regular grid. For the AGN torus and the
SPH galaxy models, with their strong density gradients, the
barycentric grids contains more cells than the regular grids, with
differences of 11\% and 2\% respectively.

To understand this, consider a subnode with
$\delta=10\,\delta_{\text{max}}$ and a strong density gradient within
the cell. In the barycentric octree, this cell will be subdivided and
each of the subnodes will contain approximately $\delta \approx
1.25\,\delta_{\text{max}}$, which means they will all be subdivided
again, resulting in 64 cells in total. In a regular grid, it is
possible that one subnode contains most of the mass and the remaining
seven subnodes do not need to subdivided again. Most probably, several
(but not all) of the children of this one subnode will have to
subdivided again. This could lead to a set of leaf cells with, on
average, a deeper level of subdivision, but the total number of cells
could be either less or more compared to the barycentric grid, subtly
depending on the distribution of the density within the cell.


The main goal of this Research Note is to find the most efficient
algorithm for the grid construction and traversal. We tested the
efficiency by doing accurate timings of the Monte Carlo routine, for
each of the 18 models in our test suite (i.e.\ for the three
geometries, the two grid subdivision methods, and the three grid
traversal algorithms).  Obtaining precise and repeatable timings is
trickier than it might seem, as even in single-thread simulations
modern hardware features can complicate matters. We performed our
timing tests on a server installed in a temperature-controlled room,
in a single execution thread on otherwise idle computers. Numbering
the 18 tests from 1 to 18, we ran the sequence 1--18, 1--18,
18--1. The timing variations between the three runs turned out to be
less than 2 seconds for each simulation, well below the differences
between the various methods.

The average run times for each of the different runs\footnote{Octree
  construction times are not considered in this table. In the current
  SKIRT implementation (which was not optimized for octree
  construction, as this needs to be done only once for each
  simulation), the octree construction time is a few minutes at most
  (depending on the complexity of the dust density field), whereas a
  typical full-scale radiative transfer simulation can last several
  hours. In other applications where the efficiency of the octree
  construction is important, more advanced tree construction
  algorithms can be applied \citep{Sundar2008, p4est}.} can be found
in the top three rows of Table~{\ref{Statistics.tab}}. Ignoring
intricacies such as loop overhead, the run time of each simulation can
schematically be written as
\begin{equation}
  t_{\text{run}}
  =
  N_{\text{pp}}  \langle N_{\text{path}}\rangle
  \left[
  \langle t_{\text{launch}}\rangle
  +
  \langle N_{\text{cross}}\rangle
  \Bigl( \langle t_{\text{id}}\rangle + \langle t_{\text{cross}}
  \rangle \Bigr)
  \right] ,
\end{equation}
where $N_{\text{pp}}$ is the total number of photon packages in the
simulation, $\langle N_{\text{path}}\rangle$ is the average number of
random straight paths in a photon package, $\langle
t_{\text{launch}}\rangle$ is the average time needed to generate the
starting location and orientation of a path, $\langle N_{\text{cross}}
\rangle$ is the average number of cells crossed by a path, $\langle
t_{\text{id}} \rangle$ is the average time necessary to identify the
next cell along the path, and $\langle t_{\text{cross}} \rangle$ is
the average time necessary to cross this cell (i.e.\ to find the exit
point and the covered pathlength within the cell). Of these six
factors, $N_{\text{pp}}$ is the same for all runs, and $\langle
t_{\text{cross}} \rangle$ is nearly identical (tiny differences can
occur because of cache misses). The quantities $\langle
N_{\text{path}}\rangle$ and $\langle t_{\text{launch}}\rangle$ are
similar for all runs corresponding to a given test model. Furthermore,
$\langle t_{\text{launch}}\rangle$ is expected to be small compared to
the grid traversal, which was indeed confirmed in separate timing
experiments. For a fixed geometry, the differences in run time are
therefor dominated by differences in $\langle N_{\text{cross}} \rangle$
and $\langle t_{\text{id}} \rangle$, where the former depends only on
the grid subdivision method (regular or barycentric), and the latter
depends on both the grid subdivision method and the grid traversal
method.

The first clear result that we find from a comparison of the run
times listed in Table~{\ref{Statistics.tab}}, is that for all
simulations in our suite, the neighbour list algorithm is the fastest
method to traverse photon packages through the dust grid. It is faster
than the other two methods, which are almost equally efficient, by
about 20\%. For the neighbour list method, $\langle
t_{\text{id}}\rangle$ is proportional to the number of neighbouring
dust cells that needs to be tested every time a cell is
crossed. In this respect, it is important to make the distinction
between the average number of neighbours of each cell wall (Row 
9 in Table~{\ref{Statistics.tab}}), the average number of neighbours of
every wall crossed (Row 10) and the number of neighbours that
need to be tested for each crossing (Row 11). These statistics
turn out to be substantially different, because some cells are crossed
more often than others and because neighbouring cells with the largest
overlap area are tested first in the neighbour search algorithm. The
ordering of the neighbour list by overlapping surface area makes the
neighbour list algorithm extremely efficient.

Another remarkable result is that simulations with a barycentric grid
are {\em{always}} slower than the corresponding simulations with a
regular grid, even if the number of cells in the barycentric grid is
smaller than in the regular grid.  At first sight, this is
surprising. Consider, for example, the top-down method, where $\langle
t_{\text{id}}\rangle$ is proportional to the average level of every
cell crossed. As the average level of each cell in a barycentric grid
is lower than in a regular grid (Rows 5 and 6 in
Table~{\ref{Statistics.tab}}), one would expect the barycentric grid
to be more efficient than the regular grid. However, the difference in $\langle
N_{\text{cross}} \rangle$, i.e.\ the average number of cells crossed
per path, means that the opposite is true. Row 8 of Table~{\ref{Statistics.tab}} shows that
$\langle N_{\text{cross}} \rangle$ is indeed systematically larger for
the barycentric octrees compared to the corresponding regular
octrees. The reason for the difference in $\langle N_{\text{cross}}
\rangle$ (and so in efficiency) is geometrical in nature. By
construction, barycentric grids are more irregular in structure, with
more neighbours per cell wall. In a regular grid, every cell
has, on average, only 1.03 neighbours per wall, whereas the barycentric
grid cells have, on average, 1.86 neighbours per wall. Having more
neighbours and a more irregular distribution also leads to shorter
paths crossed per cell and so more cells along each path.

\section{Discussion and conclusion}
\label{Conclusion.sec}

In this Research Note, we have critically investigated the use of hierarchical
octree grids to partition a dusty medium in the frame of 3D dust
radiative transfer codes, but the results are equally applicable to
other Monte Carlo transport problems. Octree grids can refine the
gridding in higher density regions without the need to create
undesirable dust cells in low density regions. We have implemented a
flexible octree structure in the 3D Monte Carlo code SKIRT
\citep{2003MNRAS.343.1081B, 2011ApJS..196...22B}, which allows for
either a regular or a barycentric iterative subdivision of the
cells. We implemented two alternative methods for octree traversal
(neighbour list search and bookkeeping), and we compared them with a
more straightforward top-down method.  We ran simulations on three
representative astrophysical models (spiral galaxy model, clumpy AGN
torus, and galaxy from SPH simulation) to test the efficiency of the
octree construction types and the traversal methods in typical Monte
Carlo radiative transfer environments.

Our main conclusions are the following:
\begin{enumerate}
\item A barycentric subdivision leads to a lower average level of
  subdivision compared to regular subdivision. This, however, does not
  directly imply that barycentric octree grids contain fewer dust
  cells than the corresponding regular grids.
\item The neighbour list method is consistently the most efficient
  way to calculate paths through an octree, with a 20\% advantage
  compared to the other two methods. The efficiency of the neighbour
  list method is achieved in part by ordering the neighbour lists
  according to decreasing overlap area.
\item Octree traversal is less efficient in barycentric octrees than
  in the corresponding regular octrees in all cases, even for
  simulations in which the regular grid contains more grid cells. The
  reason is the average number of grid cells crossed by a path, which
  is significantly larger in barycentric grids compared to the
  corresponding regular octree grids.
\end{enumerate}
Based on the above, we conclude that, while they are designed to
follow the mass distribution more closely,  barycentric octree grids
are less efficient than regular octree grids, and we strongly
recommend the neighbour list method as the preferred method to traverse
octrees in the frame of 3D radiative transfer simulations.

Taking a step back, the underlying goal of this research is to find
better ways for partitioning a dusty medium into cells in the context
of a 3D Monte Carlo RT simulation.  We have considered a criterion
based on the mass in each cell. Is
this the best possible criterion? One could consider an alternative
criterion based on the density gradient within each cell. This
may be particularly interesting for a density distribution with strong
gradients or sharp boundaries such as the AGN torus model, where the
barycentric grid subdivision results in large elongated grid cells in
regions with a sharp boundary (see Fig.~{\ref{Models.fig}}).
Furthermore, for radiative transfer simulations including thermal
emission, the ideal grid depends not only on the dust density, but
also on the mean intensity of the radiation field, which can only be
determined through the radiative transfer simulation itself. Possible
ways to deal with this include using knowledge of the source
function when constructing the dust grid \citep{2005A&A...439..153S},
or refining the grid structure based on low-resolution
pre-calculations \citep{2006A&A...456....1N}. The most
general approach isprobably to use an iterative scheme for the determination of
the grid: start with an initial grid structure derived from the dust
density field (based on total mass and/or density gradient), determine
the radiation field in every cell using the radiative transfer
simulation, and iteratively refine/redetermine the grid based on the
properties of the cell-to-cell variance of the radiation field. This
is beyond the scope of the present Research Note and may be
investigated in future work.

\begin{acknowledgement}
  We thank the referee for his/her constructive referee report that
  improved the content and presentation of this Research Note.  WS
  acknowledges the support of Al-Baath University and The Ministry of
  High Education in Syria in the form of a research grant. This work
  fits in the CHARM framework (Contemporary physical challenges in
  Heliospheric and AstRophysical Models), a phase VII Interuniversity
  Attraction Pole (IAP) programme organised by BELSPO, the BELgian
  federal Science Policy Office. MS acknowledges the support of the
  Ministry of Education, Science and Technological Development of the
  Republic of Serbia through the projects `Astrophysical Spectroscopy
  of Extragalactic Objects' (176001) and `Gravitation and the Large
  Scale Structure of the Universe' (176003). 
\end{acknowledgement}

\bibliographystyle{aa} 
\bibliography{OctreeRN}

\end{document}